# Regime of small number of photons in the cavity for a single-emitter laser

N V Larionov[1,2]

[1] Peter the Great St. Petersburg Polytechnic University, 195251, St. Petersburg, Russia
[2] St. Petersburg State Marine Technical University, 190121, St. Petersburg, Russia

E-mail: larionov.nickolay@gmail.com

**Abstract.** The simplest model of a single-emitter laser generating in the regime of small number of photons in the cavity is theoretically investigated. Based on a system of equations for different moments of the field operators the analytical expressions for mean number of photons and its dispersion are obtained. Using the master equation approach the differential equation for the phase-averaged quasi-probability $Q$ is derived. For some limiting cases the exact solutions of this equation are found.

## 1. Introduction

Nowadays, there are many experimental and theoretical studies aimed at creating specific quantum states of both atomic and field systems [1-3]. These systems, for example, can be used as sources of non-classical states of light which are in demand in such areas of physics as quantum computer science, quantum communications, quantum cryptography, and quantum frequency standards [4-9]. Various studies are aimed at creating such sources. In particular, there are experiments in which systems, consisting of only one or several quantum emitters, are used to obtain certain states of light [1,10]. The properties of single emitters, their Fermi statistics, are displayed on the state of electromagnetic radiation, which makes it possible to obtain, for example, sub-Poissonian light.

One of the fundamental models of quantum optics is the model of a single-emitter laser (SEL): a two-level atom with incoherent pumping interacting with a damped single cavity mode. This model was first considered in [11] and has since been studied by many other authors. An essential contribution to the understanding of the behavior of a SEL and related physical phenomena was made by a scientific group headed by S. Ya. Kilin from the B. I. Stepanov Institute of Physics, Belarus (see [12,13] and references there). One of the theoretical approaches used in the works of this group is based on the analysis of the equation for the density operator of the system, written for such quasi-probability distribution functions as $P$ and $Q$. These quasi-probabilities allow one to find normally- and antinormally ordered correlation functions of field operators, respectively.

In [14], for the case of stationary SEL generation, we derived a second-order linear homogeneous differential equation for the $P$-function averaged over the phase. In the limiting case, when the atom-field coupling is stronger than the coupling of the field to the reservoir that provides its decay, an approximate solution of this equation was obtained. This solution, which is unperturbed solution in the problem of a small parameter at the highest derivative (a singularly perturbed problem), for certain values of the laser parameters gives good agreement with numerical calculations and, moreover, contains some solutions previously found in [12,13]. Our further analysis of this equation allowed us to derive an approximate solution for the $P$-function, which, unlike solution in [14], could become negatively defined [15]. When the atom Fermi statistics begin to strongly impose itself on the photon statistics, the $P$-function exhibits non-classical behavior, which makes it difficult to analyze the behavior of the SEL based on mentioned equation. Therefore, there is a natural desire to get a similar equation, but for a "good" quasi-probability.

In this paper, we consider the stationary generating of a SEL in the regime of small number of photons in the cavity, accompanied by strong-coupling regime. Using the approach based on an infinite system of equations for different moments of the field operators [16,17] the analytical expressions for a mean number of photons and its dispersion are obtained. Unlike our previous paper [16], where the specific example was considered, these expressions are obtained in the more general case. Based on the master equation for density operator the equation for the phase-averaged $Q$-function for a SEL is derived. For the limiting cases in which the pumping tends to zero two simple exact solution of this equation are found.

## 2. Model of a SEL. Equation for the phase averaged $Q$-function. System of equations for different moments of the field operators

The simplest model of a SEL is represented by two-level atom interacting with a single damping cavity mode. The atom is characterized by two rate constants: $\gamma/2$ - the spontaneous emission rate from the upper atomic level $|2\rangle$ to the lower atomic level $|1\rangle$; $\Gamma/2$ - the incoherent pumping rate from the level $|1\rangle$ to the level $|2\rangle$. The cavity mode decay rate is denoted as $\kappa/2$ and the coupling constant between atom and cavity mode is denoted as $g$.

The master equation for density operator $\hat{\rho}$ is

$$\frac{\partial \hat{\rho}}{\partial t} = -\frac{i}{\hbar}[\hat{V},\hat{\rho}] + \frac{\kappa}{2}\left(2\hat{a}\hat{\rho}\hat{a}^\dagger - \hat{a}^\dagger\hat{a}\hat{\rho} - \hat{\rho}\hat{a}^\dagger\hat{a}\right) + \frac{\gamma}{2}\left(2\hat{\sigma}\hat{\rho}\hat{\sigma}^\dagger - \hat{\sigma}^\dagger\hat{\sigma}\hat{\rho} - \hat{\rho}\hat{\sigma}^\dagger\hat{\sigma}\right) + \\ + \frac{\Gamma}{2}\left(2\hat{\sigma}^\dagger\hat{\rho}\hat{\sigma} - \hat{\sigma}\hat{\sigma}^\dagger\hat{\rho} - \hat{\rho}\hat{\sigma}\hat{\sigma}^\dagger\right), \quad \hat{V} = i\hbar g\left(\hat{a}^\dagger\hat{\sigma} - \hat{\sigma}^\dagger\hat{a}\right), \quad (1)$$

where $\hat{a}^\dagger, \hat{a}$ are the photon annihilation and creation operators in the cavity mode, $\hat{\sigma} = |1\rangle\langle 2|$ ($\hat{\sigma}^\dagger = |2\rangle\langle 1|$) is the operator of polarization of the two-level atom and $\hat{V}$ is the interaction operator between atom and cavity mode.

From (1), using well-known rules, one can obtain the following system for the antinormally-ordered representation of the density matrix of our laser $\rho_{ik}(z,z^*,t) = \langle i|\hat{\rho}(z,z^*,t)|k\rangle$ over the coherent states $|z\rangle$ of the field and over the projections on the atomic states $|i\rangle, |k\rangle$, $i,k = 1,2$

$$\begin{cases} \frac{\partial Q}{\partial t} = \frac{\partial}{\partial z}\left[\frac{\kappa}{2}\left(zQ + \frac{\partial Q}{\partial z^*}\right) - g\rho_{21}\right] + \frac{\partial}{\partial z^*}\left[\frac{\kappa}{2}\left(z^*Q + \frac{\partial Q}{\partial z}\right) - g\rho_{12}\right], \\ \frac{\partial D}{\partial t} = (\Gamma - \gamma)Q - (\Gamma + \gamma)D + \frac{\partial}{\partial z}\left[\frac{\kappa}{2}zD - g\rho_{21}\right] + \frac{\partial}{\partial z^*}\left[\frac{\kappa}{2}z^*D - g\rho_{12}\right] - 2g\left(z^*\rho_{21} + z\rho_{12}\right) + \kappa\frac{\partial^2 D}{\partial z \partial z^*}, \quad (2) \\ \frac{\partial \rho_{21}}{\partial t} = -\frac{(\Gamma + \gamma)}{2}\rho_{21} + \frac{\kappa}{2}\left[\frac{\partial}{\partial z}z\rho_{21} + \frac{\partial}{\partial z^*}z^*\rho_{21}\right] + g\left[zD + \frac{1}{2}\frac{\partial}{\partial z^*}(D - Q)\right] + \kappa\frac{\partial^2 \rho_{21}}{\partial z \partial z^*}, \end{cases}$$

where we introduce the following quasi-probabilities: the $Q$-function $Q = \rho_{11} + \rho_{22}$ and the difference $D = \rho_{22} - \rho_{11}$, where $\rho_{ii} \equiv \rho_{ii}(z,z^*,t)$; the coherence $\rho_{ik} \equiv \rho_{ik}(z,z^*,t)$ for $i \neq k$. The physical meaning of the additional two quasiprobabilities $D$ and $\rho_{21} = \rho_{12}^*$ follows from theirs mean values: $\langle D\rangle \equiv \langle \hat{\sigma}_z\rangle = \int D d^2z$ - the mean value for the atomic inversion; $\langle \rho_{21}\rangle \equiv \langle \hat{\sigma}\rangle = \int \rho_{21} d^2z$ - the mean value of the atomic polarization.

The first equation in the system (2) can be written in the continuity equation form [14]: $\partial Q/\partial t + \text{div}\,\vec{J} = q$, where $\text{div} = (\partial/\partial z, \partial/\partial z^*)$ and quasi-probability current is defined as $\vec{J} = (J, J^*)$ with $J = -\kappa/2(z + \partial/\partial z^*)Q$; the source $q = -g(\partial \rho_{21}/\partial z + \partial \rho_{12}/\partial z^*)$ can be also represented as divergence of some vector.

In the stationary regime, from above continuity equation it is easy to obtain the following relation between phase-averaged $Q$-function $Q(I)$ and sum of phase-averaged coherencies $\rho_\Sigma(I) = \rho_{12}(I) + \rho_{21}(I)$

$$\rho_\Sigma(I) = \frac{\kappa}{g} I^{1/2} \left[ Q(I) + \frac{dQ(I)}{dI} \right], \quad (3)$$

where

$$Q(I) = \frac{1}{2\pi} \int_0^{2\pi} Q(I,\varphi) d\varphi, \quad \rho_{12}(I) = \rho_{21}^*(I) = \frac{1}{2\pi} \int_0^{2\pi} e^{i\varphi} \rho_{12}(I,\varphi) d\varphi$$

and we introduced the polar coordinates $I^{1/2}, \varphi$ of the complex variable $z = I^{1/2} e^{i\varphi}$.

In the same stationary regime, after averaging all over the phase, the last two equations from system (2) can be written as

$$\begin{cases} (\omega - \eta) Q(I) - (\omega + \eta) D(I) - I^{1/2} \rho_\Sigma(I) = \dfrac{d}{dI}\left[ \dfrac{1}{2} I^{1/2} \rho_\Sigma(I) - \tau I D(I) - \tau I \dfrac{dD(I)}{dI} \right], \\ (\omega + \eta) \rho_\Sigma(I) + \dfrac{\tau}{2I} \rho_\Sigma(I) - I^{1/2}\left[ 2D(I) + \dfrac{d}{dI}(D(I) - Q(I)) \right] = 2\tau \dfrac{d}{dI} I \left[ \rho_\Sigma(I) + \dfrac{d\rho_\Sigma(I)}{dI} \right], \end{cases} \quad (4)$$

where $D(I) = (1/2\pi) \int_0^{2\pi} D(I,\varphi) d\varphi$ is the phase-averaged difference.

A system of $n$ coupled differential equations, as is well known from the corresponding theory, can be reduced to one differential equation of an order higher than $n$ or equal to $n$. In our case, the system (4) together with relation (3) is equivalent to a single 5$^{\text{th}}$-order differential equation for function $Q(I)$

$$\sum_{\nu=0}^{5} f_\nu(I) Q^{(\nu)}(I) = 0,$$

$$f_5(I) = b_{02} I^2 + b_{03} I^3, \; f_4(I) = b_{11} I + b_{12} I^2 + b_{13} I^3, \; f_3(I) = b_{20} + b_{21} I + b_{22} I^2 + b_{23} I^3, \quad (5)$$

$$f_2(I) = b_{30} + b_{31} I + b_{32} I^2 + b_{33} I^3, \; f_1(I) = b_{40} + b_{41} I + b_{42} I^2, \; f_0(I) = b_{50} + b_{51} I + b_{52} I^2,$$

where $Q^{(\nu)}(I) \equiv d^\nu Q(I)/dI^\nu$ and coefficients $b_{ik} = b_{ik}(\Gamma, \gamma, \kappa, g)$ are written out in the appendix.
Note that to exclude the phase-averaged difference and its derivatives from system (4), another equation was obtained by differentiating the second equation in this system.
Equation (5) is the first main result of this section.

Next thing we need is the infinite system of coupled equations for different moments of the field operators $\langle (\hat{a}^\dagger)^k \hat{a}^k \rangle$, $0 \leq k < \infty$. This system of equations for a SEL was first derived by G. S. Agarwal and S. Dutta Gupta [17]. In [16] we introduced one simple way to obtain this system using the equation for phase-averaged $P$-function. We also showed there that in the regime of small number of photons in the cavity $\langle n \rangle \equiv \langle \hat{a}^\dagger \hat{a} \rangle \approx 1$, when $\Gamma \approx \kappa$, our laser can be adequately described by the first three reduced equations from this infinite system. Reducing of these equations means retaining only the terms arising from moments $\langle (\hat{a}^\dagger)^k \hat{a}^k \rangle$ with $0 \leq k \leq 3$. Here, without detailed derivation, we present this finite system of equations and its solutions, obtained using the Cramer's rule

$$\begin{cases} c_{11} \langle n \rangle + c_{12} \langle n^2 \rangle = B, \\ c_{21} \langle n \rangle + c_{22} \langle n^2 \rangle + c_{23} \langle n^3 \rangle = a_{10}, \\ c_{31} \langle n \rangle + c_{32} \langle n^2 \rangle + c_{33} \langle n^3 \rangle = 0, \end{cases} \begin{cases} \langle n \rangle = \dfrac{B(c_{22} c_{33} - c_{23} c_{32}) - a_{10} c_{33}}{\Delta}, \\ \langle n^2 \rangle = \dfrac{B(c_{23} c_{31} - c_{21} c_{33}) + a_{10} c_{11} c_{33}}{\Delta}, \\ \langle n^3 \rangle = \dfrac{B(c_{21} c_{32} - c_{22} c_{31}) + a_{10}(c_{31} - c_{11} c_{32})}{\Delta}, \end{cases} \quad (6)$$

where coefficients $c_{ik}, a_{10}, B$ are also taken out at the appendix and $\Delta$ is a determinant.

Expressions for $\langle n \rangle, \langle n^2 \rangle, \langle n^3 \rangle$ (6) are the second main result in this section.

In the next section results (5), (6) will be used for analyzing the statistical properties of a SEL.

## 3. Results

From now on, the following three dimensionless constants will be used: $\omega = \Gamma/2g$, $\eta = \gamma/2g$, $\tau = \kappa/2g$. Using results (6) we plot the mean number of photons in the cavity mode $\langle n \rangle$ and the Mandel Q-parameter $Q = (\langle n^2 \rangle - \langle n \rangle^2)/\langle n \rangle - 1$ as functions of $\omega$ (see figure). Because of condition $\omega \approx \tau$ we consider two special cases: $\omega = \tau$, $\omega = \tau/2$, i.e. the dimensionless cavity decay rate $\tau$ is changed together with the dimensionless pumping rate $\omega$.

The obtained dependences of $\langle n \rangle$ and $Q$ on pumping rate is well known: self-quenching effect when $\omega \to \infty$; characteristic peak for the Mandel Q-parameter indicates on transition from regime when the field statistics is determined by atomic spontaneous relaxation out of the cavity mode to the regime when the atom Fermi statistics starts to show up better; the negative value of the Mandel Q-parameter corresponds to photon antibunching effect which manifest itself better for strong-coupling regime ($\eta \ll 1$) and for certain ratio between $\omega$ and $\tau$. If we equate $\eta = 0$ we get the results, obtained in [16].

From presented calculation one can see that analytical results are in a very good agreement with numerical simulation of the master equation (we presented numerical simulation only for the case $\eta = 0.5$, for other case $\eta = 0.1$ we have the same excellent coincidence between analytical and numerical results). Thus, the assumption was confirmed that for a complete description of the statistical properties of a SEL in the regime when $\omega \approx \tau$, it is sufficient to limit ourselves to the system of equations for the field moments $\left\langle \left(\hat{a}^\dagger\right)^k \hat{a}^k \right\rangle$ with $k < 4$ [16].

Let us use obtained equation for phase-averaged Q-function (5) to find analytical results for two simple limiting cases: 1) $\eta \neq 0$, $\omega = \tau \to 0$ and 2) $\eta = 0$, $\omega = \tau \to 0$. In the first and second cases from (5) we obtain the following simple equations

$$1)\ Q^{(1)}(I) + Q(I) = 0,\ 2)\ 2I(1-I)Q^{(2)}(I) + (3 + 3I - 4I^2)Q^{(1)}(I) + (1 + 2I - 2I^2)Q(I) = 0, \quad (7)$$

and the corresponding solutions are

$$1)\ Q_1(I) = e^{-I},\ 2)\ Q_2(I) = e^{-I} \frac{\cosh\left[(2I)^{1/2}\right] + (2I)^{-1/2} \sinh\left[(2I)^{1/2}\right]}{1 + (2e\pi)^{1/2} \operatorname{erf}(2^{-1/2})}. \quad (8)$$

Solution for the first case is the well-known Q-function for a vacuum field state: $\langle n \rangle = \int_0^\infty Q_1(I) I dI - 1 = 0$. From solution for the second case one can obtain the same results for $\langle n \rangle$ and $\langle n^2 \rangle$ as in [16]: $\langle n \rangle = \int_0^\infty Q_2(I) I dI - 1 = 0.630843$, $\langle n^2 \rangle = \int_0^\infty Q_2(I) I^2 dI - 3\langle n \rangle - 2 = 1$.

Note that the function $Q_2(I)$ can be easily obtained using the phase-averaged P-function $P(I)$ derived in [16]

$$P(I) = C_0 \frac{I \cdot e^I}{(1-2I)^{3/2}},\ Q_2(I) = \int_0^\infty P(I') e^{-(I'+I)} I_0\left[2(I'I)^{1/2}\right] dI', \quad (9)$$

where $I_0[x]$ is the modified Bessel function of the first kind, $C_0$ is the normalization constant and we use the well-known rules for relation between $Q(I)$ and $P(I)$ distribution functions. Due to the singularity of the P-function at a point $I = 1/2$ the Cauchy principal value should be defined for the integral (9).

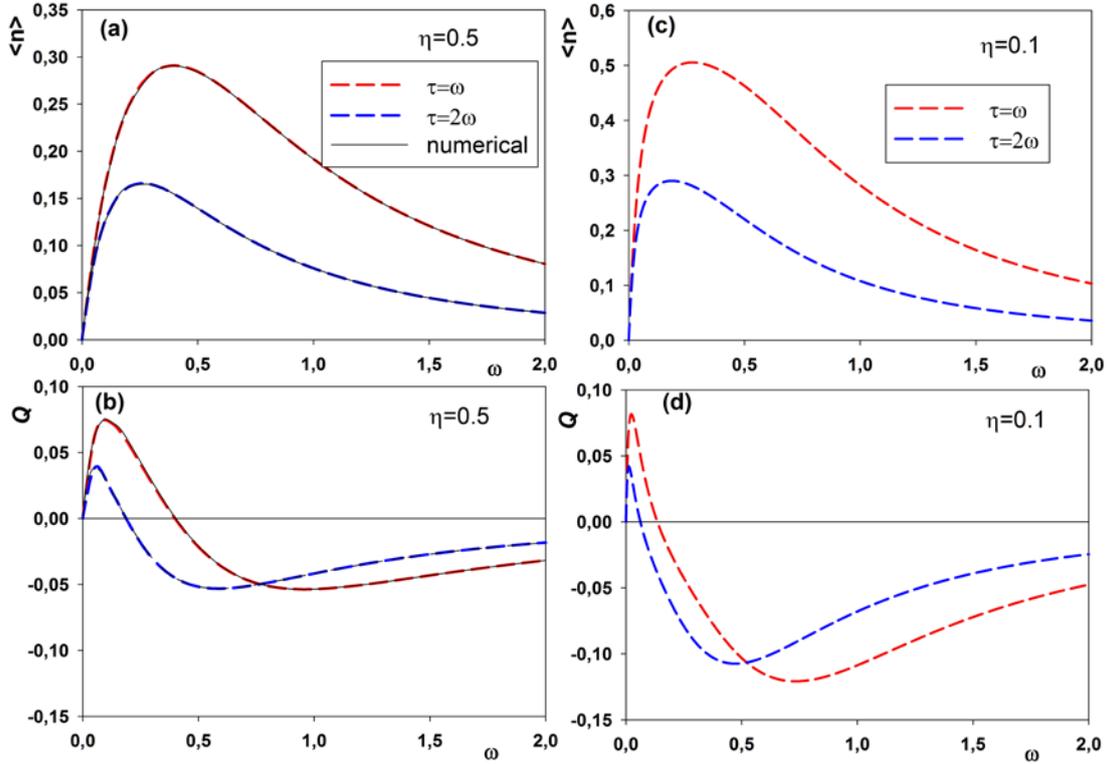

**Figure.** The mean number of photons in the cavity mode (a), (c) and corresponding Mandel $Q$-parameter (b), (d) vs $\omega$. Solid black lines (a), (b) – numerical calculation of (1); Dashed colour lines – analytical results obtained with the help of (6); $\tau = \omega$ – dashed red lines, $\tau = 2\omega$ – dashed blue lines.

## 4. Conclusion

In this paper, we considered the simple model of a single-emitter laser generating in the strong coupling regime $\eta < 1$, accompanied by the regime of small number of photons in the cavity mode $\omega \approx \tau$. Based on a system of coupled equations for certain moments of the field operators $\left\langle \left(\hat{a}^\dagger\right)^k \hat{a}^k \right\rangle$ with $k < 4$, derived in [16], the analytical expressions for the mean number of photons and its variance was obtained. These analytical results demonstrate excellent agreement with the results of numerical simulation of the master equation.

For the laser under consideration, the stationary differential equation for the phase-averaged $Q$-function $Q(I)$ was derived from the master equation. In the two limiting cases 1) $\eta \neq 0$, $\omega = \tau \to 0$ and 2) $\eta = 0$, $\omega = \tau \to 0$ this differential equation was resolved and clear analytical expressions was obtained. In the first limiting case we obtained the explicit result – the $Q$-function for the vacuum state of the cavity mode. The expression for the $Q$-function obtained in the second case is in agreement with the result from [16], where this problem was considered with the help of phase-averaged $P$-function.

The derived equation for $Q(I)$ is a fifth-order differential equation and that is the problem. The corresponding equation for phase-averaged $P$-function [14] is a second-order differential equation. The high order of the differential equation for $Q(I)$ is the price we pay to get away from the problems inherent in the $P$-function.

## Acknowledgments

The work was supported by the State program for the fundamental research (theme code FSEG-2020-0024).

# Appendix

The coefficients in the equation (5)

$$b_{02} = -2\tau^3(\omega+\eta+\tau), b_{03} = 4\tau^4; b_{11} = -12\tau^3(\omega+\eta+\tau), b_{12} = 2\tau^3(7\tau-3\omega-3\eta), b_{13} = 12\tau^4;$$

$$b_{20} = -12\tau^3(\omega+\eta+\tau), b_{21} = -\tau^2(26\eta\tau - 3\eta^2 + 21\tau^2 - 6\eta\omega + 26\tau\omega - 3\omega^2), b_{22} = 12\tau^3(4\tau-\omega-\eta),$$

$$b_{23} = 12\tau^4, b_{30} = -2\tau^2(8\eta\tau - 3\eta^2 + 15\tau^2 - 6\eta\omega + 8\tau\omega - 3\omega^2);$$

$$b_{31} = -2\tau(\eta+\tau-3\eta^2\tau+13\eta\tau^2+\omega-6\eta\tau\omega+13\tau^2\omega-3\tau\omega^2), b_{32} = 2\tau^2(2-7\eta\tau+23\tau^2-7\tau\omega), b_{33} = 4\tau^4;$$

$$b_{40} = -\eta^3\tau+\eta^2(8\tau^2-1-3\tau\omega)-\tau(3\tau+24\tau^3+3\omega-\tau^2\omega-8\tau\omega^2+\omega^3)+\eta(\tau^3-\omega+16\tau^2\omega-\tau(4+3\omega^2)),$$

$$b_{41} = \tau(5\eta^2\tau+15\tau^3-4\omega-20\tau^2\omega-2\eta(1+10\tau^2-5\tau\omega)+\tau(5\omega^2-2)), b_{42} = 2\tau^2(4-3\eta\tau+7\tau^2-3\tau\omega);$$

$$b_{50} = -\eta^3\tau-6\tau^4+5\tau^3\omega+\omega^2-\tau\omega^3+\eta^2(2\tau^2-1-3\tau\omega)+\eta\tau(5\tau^2-4+4\tau\omega-3\omega^2)+\tau^2(2\omega^2-3),$$

$$b_{51} = 2\tau(\eta^2\tau+3\tau^3-2\omega-4\tau^2\omega+\tau\omega^2+2\eta\tau(\omega-2\tau)), b_{52} = 4\tau^2.$$

The coefficients in the system (6)

$$c_{11} \equiv A = \frac{\omega+\eta+\tau}{2(\omega+\eta)} - \left(\frac{\omega-\eta+\tau}{2\tau} - \frac{(\omega+\eta+\tau)^2}{2}\right), c_{12} = 1, B = \frac{\omega}{2\tau}\frac{(\omega+\eta+\tau)}{(\omega+\eta)};$$

$$c_{21} = (6a_{02} - 12a_{03} - 2a_{11} + 3a_{12} + a_{20} - a_{21} + 2a_{22}), c_{22} = (12a_{03} - 3a_{12} + a_{21} - 3a_{22}), c_{23} = a_{22};$$

$$c_{31} = (40a_{03} + 3a_{11} - 8a_{12} - a_{20} + 2a_{21} - 12a_{02} - 2a_{10}), c_{32} = (-60a_{03} - 3a_{11} + 12a_{12} + a_{20} - 3a_{21} + 12a_{02}),$$

$$c_{33} = (20a_{03} - 4a_{12} + a_{21});$$

$$a_{02} = \frac{\tau^3}{2}(\tau-\omega-\eta), a_{03} = \tau^4, a_{10} = \frac{\omega}{4}(\tau-\omega-\eta), a_{11} = \frac{\tau}{4}[3\eta^2\tau+9\tau^3+4\omega-12\tau^2\omega+\eta(2-12\tau^2+6\tau\omega)+$$

$$+\tau(3\omega^2-2)], a_{12} = \frac{\tau^2}{2}(7\tau^2-3\tau\eta-3\tau\omega-2), a_{20} = \frac{1}{4}[6\tau^4+\omega^2-\eta^3\tau-11\tau^3\omega-\tau\omega^3+\eta^2(6\tau^2-3\tau\omega-1)+$$

$$+\eta\tau(12\tau\omega+4-11\tau^2-3\omega^2)+\tau^2(6\omega^2-3)], a_{21} = \frac{\tau}{2}[\eta^2\tau+3\tau^3-2\omega-4\tau^2\omega+\tau\omega^2+2\eta\tau(\omega-2\tau)], a_{22} = \tau^2.$$